\documentclass[%
amsmath,%
amssymb,%
12pt,
final,
pra
]{revtex4}

\usepackage{dcolumn}
\usepackage{setspace}
\usepackage{url}
\usepackage{graphicx}
\usepackage{bm}

\newcommand{\eqn}[1]{Eq.\,(\ref{#1})}

\newcommand{\noeqn}[1]{(\ref{#1})}
\newcommand{\fig}[1]{Fig.\,\ref{#1}}

\newcommand{\tab}[1]{Tbl.\,\ref{#1}}

\newcommand{\hide}[1]{}

\begin{document}

\title{%
  Descent curves and the phases of collapse of WTC 7
}
\author{Charles M. BECK}
\email{beck.charles_m@yahoo.com}


\date{\today}

\begin{abstract}
  We examine four WTC 7 descent curves, labeled ``C,'' ``E,'' ``N,'' and ``O,''
  either anonymously published, or confidentially communicated to us.
  Descent curve describes apparent height of a collapsing building
  as a function of time.
  While all sets are mutually consistent, it is set ``C''  which suggests
  that there are three active phases of collapse.
  Phase I is a free fall for the first $H_1\simeq28$~m or $T_1\simeq2.3$~s,
  during which the acceleration $a$ is that of the gravity, $a=g=9.8$~m/s$^2$.
  In Phase II, which continues until drop $H_2\simeq68$~m,
  or $T_2\simeq3.8$~s, the acceleration is $a\simeq5$~m/s$^2$,
  while in Phase III which continues for the remaining of the data set,
  $a \simeq -1$~m/s$^2$.
  \\
  We propose that the collapse of WTC 7 is initiated by a total and sudden
  annihilation of the base (section of the building from the ground level to
  $H_1$), which then allows the top section (building above $H_1$) to
  free fall during Phase I, and then collide with the ground in Phase II
  and III.
  We interpret the latter two phases of the collapse as the top section being
  comprised of two zones, the 60\% damaged
  primary zone (below $H_2$) and the intact secondary zone (above $H_2$).
  We derive a physical model for collision of the building with the ground,
  in which we correct the ``crush-up'' model of Ba\v{z}ant and Verdure,
  J. Engr. Mech. ASCE, {\bf 133} (2006) 308.
  The magnitude of resistive force in the two zones of the top section
  obviates the catastrophic failure mechanisms of Ba\v{z}ant and
  Verdure ({\em ibid.}), and of Seffen, J. Engr. Mech. ASCE, {\bf 134} (2008) 125.
  The total duration of the collapse, assuming that Phase III continues
  to the end, is in the range $7.8-8.6$~s.
  \\
  We compare our findings to those of NIST investigators and find an agreement
  with respect to the distribution of damage in the primary zone.
  We conclude that the building was destroyed in a highly controlled
  fashion.
\end{abstract}

\maketitle
\newpage

\section{Introduction}

\begin{figure}[tph]
  \singlespacing
  \centering
  \includegraphics[scale=0.7,clip]{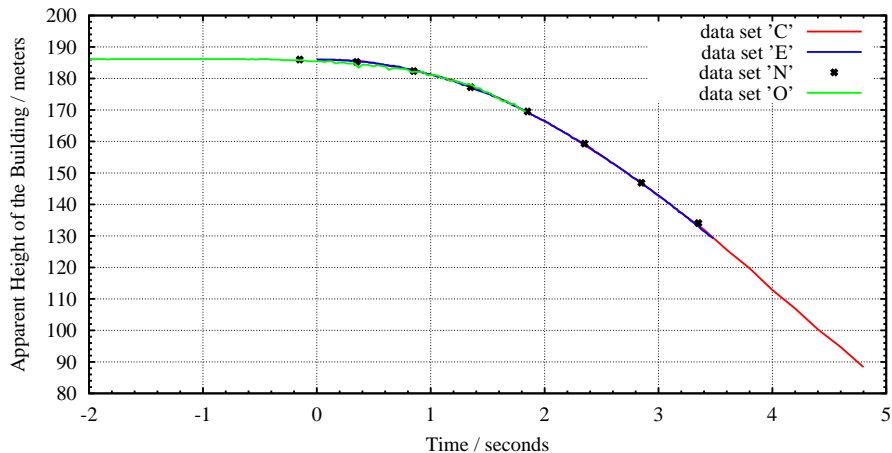}
  \caption{%
    \label{fig:1}
    Position of the top of WTC 7 as a function of time (descent curve)
    as recorded in four data sets, labeled ``C,'' ``E,'' ``N,'' and ``O.''
    Data set ``O'' covers some 1.8~s of collapse, however, it extends
    to 9 seconds before the collapse and so shows that during that time
    the building did not noticeably move.
    Data sets ``E'' and ``N'' cover the first 3.5 seconds of collapse with
    different sampling rates.
    Most of the report deals with ``C'' as it is the most extensive: it
    covers 4.8 seconds of collapse.
}
\end{figure}

World Trade Center (WTC) 7 perished after WTC 1 and 2 collapsed on September 11,
2001.
Its demise was examined by Federal Emergency Management Agency (FEMA), which
in 2002 issued a report.~\cite{fema:2002}
In the report it was claimed that the most likely cause of the collapse was the
gradual weakening of vertical columns in the lower part of the building following
their exposure to the mechanical and thermal stress.
As the sources of mechanical stress the falling debris and the earthquake from the
collapse of WTC 1 and 2 were given,
while the fires raging inside the building, some of which were fueled by the heating oil
known to have been stored in the building, were cited as a source of thermal stress.
In November 20, 2008 the NIST investigators issued a report in which
it was claimed that the fires that followed the impact of debris from the collapse of
WTC 1 (the north tower) led to the collapse of WTC 7.~\cite{nist:2008:wtc7}

One easy-to-capture feature of a collapse of any high rise building is
the motion of its top as a function of time - its descent curve.
If the falling building collides with the Earth at a relatively well defined
collision plane, then, on one hand side, the motion of the top reflects the motion
of the moving part of the building.
On the other, this feature allows us to write down the equation of motion for the
moving part.
The physical model that follows from the equation of motion, can be fitted
to the descent curve, which in turn allows us to estimate how the
collapse was initiated and when, and what was the damage distribution in
the building.

The goal of this report is the analysis of motion of the top of WTC 7
as recorded in four descent curves: ``C''\cite{wtc7:c}, ``E''\cite{wtc7:e},
``N''\cite{wtc7:dropcurve}, and ``O''\cite{wtc7:o}, cf.~\fig{fig:1}.
We develop a physical model of the descent and fit it to the curves
in order to obtain a local building's strength.
We compare so obtained values to the estimates provided by us~\cite{beck2006},
by Ba\v{z}ant and Verdure~\cite{bazant2006} and by Seffen~\cite{seffen2008}.
We propose a scenario of collapse initiation which is consistent with the
descent curve in its entirety, and which distribution of damage
prior to collapse agrees with that of the NIST investigators.~\cite{nist:2008:wtc7}

\section{Descent Curve and Its Physical Model}

\subsection{Finite Differences Analysis of Descent Curve ``N''}

\begin{table}[h]
  \caption{
    \label{tab:n:fea}
    Descent curve ``N''\cite{wtc7:dropcurve} and its acceleration as found by
    using the finite differences.
    Shown are the results for the relevant velocities:
    the mean velocities on the intervals ($\bar v_i$ for interval
    $[t_{i-1},t_i]$) and
    the momentary velocities at the ends of the intervals,
    $v_i = v(t_i)$.
    As can be seen the average acceleration $\bar a_i$ for the
    first $T_1\sim2.3$~s, or the displacement of $H_1\sim26$~m, oscillates around
    the gravity, $g=9.81$~m/s$^2$,
    indicating a free fall.
    After time $T_1$ the acceleration drops to $\sim 5$~m/s$^2$.
  }
  \begin{tabular}{|c|c|c|c|c||c|c|c||}
    \hline\hline
    index & frame & time  & displacement & $\bar v_i$ & $v_i$ & $\bar a_i$\\
    (i)   &       & (sec) & (m)          & (m/s)      & (m/s) & (m/s$^2$)\\
    \hline\hline
      0 & - &0.00 &  0.00 &      &0.00&\\
      1 & 2 &0.35 &  0.73 & 11.35&3.64&10.39\\
      2 & 3 &0.85 &  3.66 &  8.15&8.05&8.82\\
      3 & 4 &1.35 &  8.78 &  9.51&12.80&9.51\\
      4 & 5 &1.85 & 16.46 & 10.24&17.92&10.24\\
      5 & 6 &2.35 & 26.70 &  9.51&22.68&9.51\\
      \hline
      6 & 7 &2.85 & 39.14 &  5.12&25.24&5.12\\
      7 & 8 &3.35 & 51.94 &      &&\\
    \hline\hline
  \end{tabular}
\end{table}

The descent curve shows the motion of the top of the building
as a function of time.
We use the term ``top section'' to label the part of the building
which motion is common to that described by the descent curve.
Our analysis of descent of WTC 7 then has two goals:
to find the extent of the top section, and to determine
its acceleration.
We recall that the acceleration of the top section is a result
of the forces acting on it.
We discuss the forces acting on the top section later when we derive
a physical model of the event.

We start with data set ``N'' because it is short: it contains 8 points at 0.5~s apart.
A reader interested in full presentation of how the data was obtained
is kindly directed to the original publication~\cite{wtc7:dropcurve}.
We note that the estimated error of the distance is $\pm0.5$~m.

The results of the finite differences analysis of the descent curve ``N''
are given in \tab{tab:n:fea} together with two intermediate velocities:
the mean velocities $\bar v_i$ on intervals $[t_{i-1},t_i]$,
and the momentary velocities $v_i$ at the end points $t_i$.
We immediately notice that the mean acceleration of the top section
has two distinctive values, which we associate with the phases of descent
and label with Roman numerals.
During Phase I the average acceleration is $\bar a = 9.69$~m/s$^2$,
which is within 1\% from the free fall acceleration given by the
gravity $g=9.81$~m/s$^2$.
This phase lasts for the first $2.35-2.85$~s.
Phase I is thus a free fall phase.
Phase II continues during which the mean acceleration, $\bar a$,
drops to $a\sim5.1$~m/s$^2$.
This phase presumably continues for the remaining one second of recorded data.

Now that we have determined the acceleration of the top section,
we answer the question of how far below the top of the building
does the top section extend.
We notice that the 2.35~s long free fall corresponds to a distance
of $\sim26$~m.
We label this 26~m section of the building in the path of the top
section ``the base.''
The question thus becomes at what height is the bottom of ``the base.''
The clue about where is ``the base'' comes from the NIST investigation
on the collapse of WTC 1 and 2, where it is explicitely
hypothesized~\cite{nist:2005} (p.146, Sec. 6.14.4)
\begin{quote}
  \em The structure below the level of collapse initiation offered minimal resistance
  to the falling building mass at and above the impact zone.
  The potential energy released by the downward movement of the large building mass
  far exceeded the capacity of the intact structure below to absorb that
  through energy of deformation.
\end{quote}
In other words, when a steel frame (the top section) collides with
another steel frame (``the base'') it feels minimal to no resistance.
However, as indicated by the descent curve, when the top section finishes
destroying ``the base'' its acceleration suddenly drops to half its value.
We conclude that at that point the top section collides with the only
other object a falling body can collide with - the Earth's surface.
This puts ``the base'' to the base of the building, from the ground level
to the height $H_1\sim26$~m, and the top section from $H_1$ all the way
to the top.
This positioning of ``the base'' is consistent with the visual
appearance of the collapse, where the visible part of the
building moves uniformly.
Behavior of the base of the building, on the other hand, is completely
hidden in the cloud of dust created at the onset of collapse.

Here however, the following comment is due.
The visual appearance of collapse, and positioning of ``the base''
to the base of the building do not imply that the NIST hypothesis stated above
is correct.
Had the top section started its motion from $H_1\sim26$~m down and accreted
the mass of the base in its path this would create an inertial brake
which would prevent the top section from ever achieving the free
fall acceleration.
As the top section starts immediately with the free fall acceleration
this implies that $(i)$, the base is converted into a free falling rubble
prior to the top section falling through it, and $(ii)$,
the crushing of the base is not done by the top section.
We return to this point later in the report, when we discuss it in
more quantitative fashion.

Based on this argument we assert that the collapse of WTC 7 begins
with a sudden and total annihilation of the base, which, on one hand,
allows the top section to free fall for the height of the base,
and on the other, leads to the observed change in acceleration once
the top section reaches the ground.
We label euphemistically the collapse initiation moment a ``release,'' and
the height $H_1\sim26$~m from which it occurs a ``release point.''
While Phase I describes the free fall of the top section to the ground
following its release, in Phase II the top section collides with the ground.
Ba\v{z}ant and Verdure~\cite{bazant2006} label Phase II of the collapse
a ``crush-up,'' and we adhere to their terminology for the rest of
the report.

\subsection{Physical Model of Collision of a Building with the Ground}

\begin{figure}[tph]
  \singlespacing
  \centering
  \includegraphics[scale=0.5,clip]{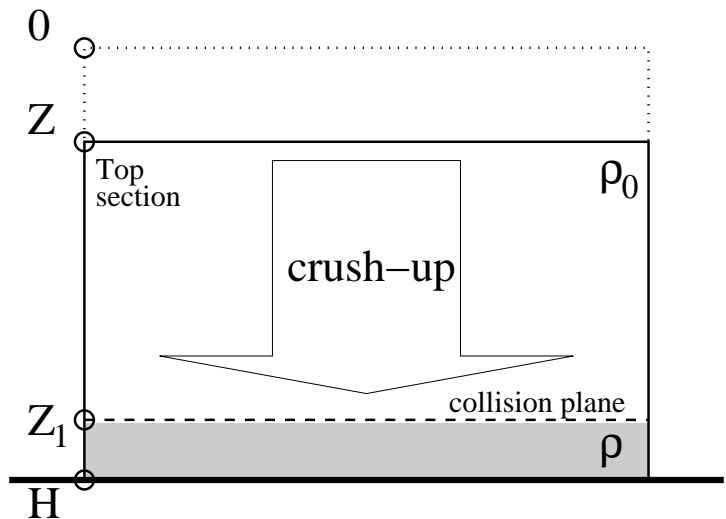}
  \caption{%
    \label{fig:2}
    ``Crush-up'' model of collapse.
    The top section, between drops $Z$ and $Z_1$, of the building
    collides with the ground at a well defined collision plane at $Z_1$.
    In the ``crush-up'' the mass and the momentum of the top section
    are lost by being transferred to the ground.
    Finite linear density of the crushed building, $\rho>0$, leads to
    a slow upward crawl of the collision plane so that $Z_1$ is not
    exactly at the ground level.
    A resistance the top section offers to its destruction at the collision
    plane is described by a resistive, or ``crushing,'' force.
}
\end{figure}

\subsubsection{Derivation of General Model}

Use of an one-dimensional model for description of a collapsing building
is justified providing that the building of interest, WTC 7, collapsed almost
perfectly in its footprint.
This remarkable feature allows us to exclude any transverse coordinates from
the analysis and use only the height to describe the motion of the building.
In doing so, we in effect average the behavior of the building
with respect to the excluded coordinates.
Most importantly, we do not need to consider a failure
of individual load-bearing structural elements, e.g., vertical columns,
and instead concentrate on their collective response.

The collapse dynamics of the building in ``crush-up'' mode
is shown in~\fig{fig:2}.
The falling building, the top of which is at $Z$, collides with
the ground at a collision plane at $Z_1$.
As a result of a finite compaction ratio the collision plane crawls up,
toward the top section.
First, we assume that the building is of uniform mass density
$\rho_0 = M/H$, where $M$ is the total mass of the building
and $H$ its height.
We introduce a compaction ratio $\kappa$, as
\begin{equation}
  \label{eq:kappa}
  \kappa = \frac{\rho_0}{\rho} \ll 1,
\end{equation}
where $\rho$ is the density of compacted building.
For simplicity, we assume that the compaction is uniform as well,
i.e., that $\rho$ is a constant.
Second, we introduce two coordinates to mark the progression
of ``crush=up,'' an apparent drop of the top of the building, $Z$, and,
a position of the collision plane, $Z_1$.
The two coordinates are connected by the requirement that the mass of the building is
conserved,
\begin{equation}
  \label{eq:massconservation}
  H \, \rho_0 = (Z_1 - Z) \, \rho_0 + (H-Z_1) \, \rho,
\end{equation}
yielding
\begin{subequations}
\begin{equation}
  Z_1 = H - \frac{\kappa}{1-\kappa} Z,
\end{equation}
\begin{equation}
  Z_1 - Z = H - \frac{1}{1-\kappa} Z,
\end{equation}
\end{subequations}
and
\begin{equation}
  \dot Z_1 = - \frac{\kappa}{1-\kappa} \dot Z.
\end{equation}

We proceed with the derivation of equation of motion for
the apparent drop of the building $Z=Z(t)$.
This is most easily accomplished by using the energy formalism.
The kinetic energy of the moving part of the building is
\begin{equation}
  \label{eq:ke}
  K(Z, \dot Z) = \frac 1 2 \, \rho_0 \, (Z_1 - Z) \, \dot Z^2 =
  \frac 1 2 \, \rho_0 \, (H - \frac{1}{1-\kappa} Z) \, \dot Z^2,
\end{equation}
while its momentum is
\begin{equation}
  \label{eq:p}
  P = \frac{\partial K}{\partial \dot Z} =
  \rho_0 \, (H - \frac{1}{1-\kappa} Z) \, \dot Z.
\end{equation}
The potential energy of the building is given by
$U(Z)=-\int_Z^{H}\ dX\ \rho(X) \, g\, X$,
giving
\begin{equation}
  \label{eq:pe}
  U(Z) = -\frac 1 2 \, \rho \, g \, \left( H^2 - Z_1^2 \right)
  -\frac 1 2 \rho_0 \, g \, \left( Z_1^2 - Z^2 \right) =
  -\frac 1 2 \rho_0 \, g \, \left( H^2 + 2 \, H \, Z - \frac{Z^2}{1-\kappa} \right).
\end{equation}
The gravitational force with respect to the coordinate $Z$ follows
from $G = -\partial U / \partial Z$, and is given by
\begin{equation}
  \label{eq:G}
  G = \rho_0 \cdot g \cdot \left( H - \frac{1}{1-\kappa} Z \right).
\end{equation}
Last energy in the problem is the ``latent'' energy $L$, from which
the force with which the building resists its destruction, call it resistive
force $R$, is derived.
We have,
\begin{equation}
  \label{eq:le}
  L = -\int_0^{Z_1 - Z} \,dX \, R(X) = L(Z_1 - Z).
\end{equation}
The resistive force $R$ with respect to the coordinate $Z$ follows,
as before, $R \equiv -\partial L / \partial Z$, and is given by
\begin{equation}
  \label{eq:R:L}
  R = -\frac{\partial L(Z_1 - Z)}{\partial Z} =
  \frac{\partial L(Z_1 - Z)}{\partial (Z_1 - Z)} \cdot
  \frac{\partial (Z_1 - Z)}{\partial Z} =
  \frac{1}{1-\kappa} \cdot R\left( H - \frac 1 {1-\kappa} Z\right).
\end{equation}

This said, the equation of motion for $Z$ follows from Newton's law,
\begin{equation}
  \label{eq:eom}
  \dot P = G + R +
  \left( \dot P \right)_{loss}.
\end{equation}
The loss of momentum (mass, energy) occurs at the avalanche front
where the momentum is transferred to the stationary part of the building.
The loss rate is $\dot Z \cdot \dot m$, where $m$ is the
mass of the moving part, yielding for the equation of motion,
\begin{equation}
  \label{eq:eom:final:1}
  \ddot Z = g + \frac{1}{1-\kappa} \cdot
  \frac {R\left( H - \frac 1 {1-\kappa} Z\right)}
    {\rho_0 \, (H - \frac{1}{1-\kappa} Z)}.
\end{equation}
We observe that while in the limit $\kappa\rightarrow0$ \eqn{eq:eom:final:1}
coincides with the result of Ba\v{z}ant and
Verdure~\cite{bazant2006}, for $\kappa\neq0$ their model
does not correctly incorporates compaction.

The resistive force $R$ describes how the building resists its destruction
at the avalanche front.
It is a function of strength of the structural elements of the building,
as well as their failure mode.
Most notable contribution comes from the vertical columns,
the strength of which varies with height $Z$.
For simplicity, we assume that the dependence of $R$ on $Z$ is
at most linear, yielding
\begin{equation}
  \label{eq:R}
  -\frac {R(Z)}{\rho_0 \, H} = g \cdot \left( r + s \, \frac Z H \right),
\end{equation}
where $r$ and $s$ are two dimensionless parameters.
With this parameterization of $R$ we obtain the ordinary differential
equation (ODE) for $Z$,
\begin{equation}
  \label{eq:eom:fema}
  \ddot Z = g \cdot (1 - \frac{s}{1-\kappa})
  - \frac {g \cdot r}{(1-\kappa) \, (1 - \frac{1}{1-\kappa} \frac Z H)}.
\end{equation}
Finally, we note that if we set $r=s=0$, we immediately obtain
the equation of motion for a free fall,
\begin{equation}
  \label{eq:freefallphase}
  {\ddot Z} = g.
\end{equation}
It has to be kept in mind that \eqn{eq:eom:fema} contains an assumption
of what happened to the base at the onset of collapse:
it was instantaneously converted into a free floating pile of debris,
which for $t>0$ starts to free fall to the ground and contributes to
motion of the collision plane at $Z_1$.

When we solve \eqn{eq:eom:fema} later, we always assume that the top
section starts its motion from rest, $Z(0) = \dot Z(0) = 0$.

Parameters of the building that enter \eqn{eq:eom:fema} are the total
height of the building, $H = 186$~m, and the parameters $r$ and $s$
of the local resistive force, $R/(M\,g)$.
Furthermore, in our simplified model the building comprises 47 floors,
each 3.66~m high, and the 14~m  high lobby.

Though it might not be obvious, the compaction parameter $\kappa$ is of
secondary importance.
Thus, in what follows we further simplify \eqn{eq:eom:fema} by taking
$\kappa\equiv0$.

\subsubsection{Apparent Weight of the Building During the Collapse}

An apparent weight the top section exerts during the collapse on
the Earth's crust, $W'/(M\,g)$, is given by
\begin{equation}
  \label{eq:weight}
  \frac{W'}{M\,g} = \left\{
  \begin{array}{lcl}
    \frac{M'}{M}, & \mbox{for} & t < 0,\\
    0,            & \mbox{for} & t \in \left[0, T_1\right>,\\
    z-\frac{H_1}H + \frac12 \, \dot z^2 + \left(r_j + s_j \cdot (1-z) \right)
      & \mbox{for} & t \ge T_1.
  \end{array}
  \right.
\end{equation}
where $z = Z/H$ is the scaled drop, and $\dot z = \dot Z / (H/T_0)$ is
the scaled velocity, with $T_0^2 = 2\,H /g$ being a free fall time from
height $H$.
Index $j$ in $r_j$ and $s_j$ keeps track of the phase of
collapse so it is a function of the time, as well.

In~\eqn{eq:weight}, $M'/M=1-H_1/H$ is the mass of the top section,
$T_1\simeq2.3$~s is the duration Phase I, the free fall, while
$H_1\sim26$~m is the distance.
The terms appearing in $W'/(M\,g)$ are, from left to right,
$(i)$, the weight of the top section that has already reached
the ground, $\rho \, g \, (Z - H_1)$;
$(ii)$, reaction force due to the change of momentum of the crushed
material at the collision plane, $\rho \dot Z^2$;
and $(iii)$, the resistive force at the collision plane, $R=R(H-Z)$.
The last term is present because the crushing of the building at the
collision plane is performed between the ground and the top section.

We believe that $W'$, and in particular its time derivative, can be used
in interpretation of the seismic signal of the building's collapse.
As an attempt to connect the two brings forth numerous
additional complications which need to be properly addressed,
we leave this topic to future publications.

\section{Descent Curve ``C''}

\begin{figure}[tph]
  \singlespacing
  \centering
  \includegraphics[scale=0.9,clip]{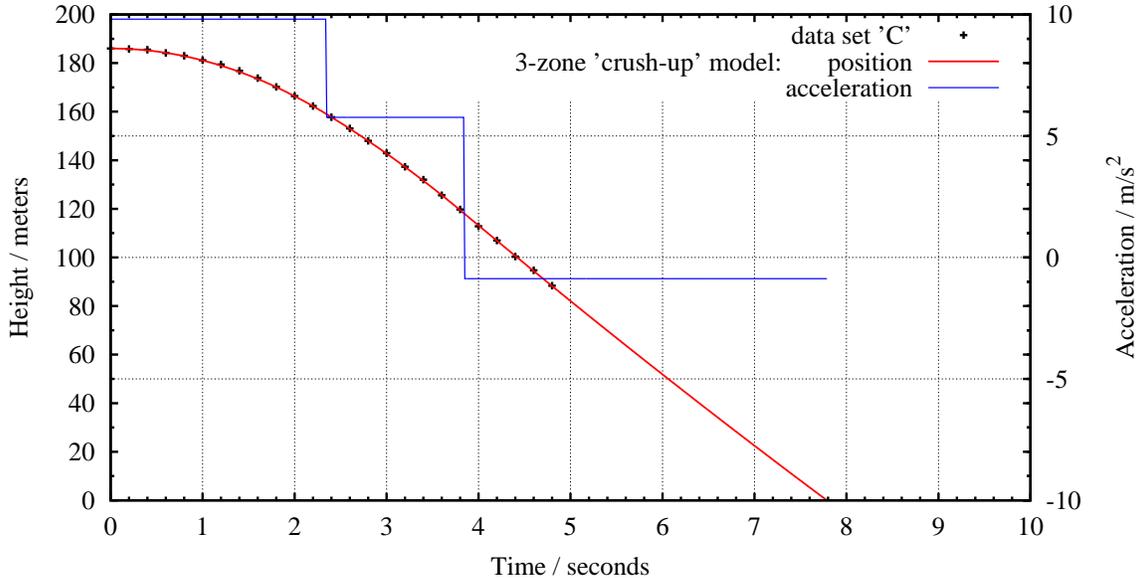}
  \caption{%
    \label{fig:3}
    Results of fitting to data set ``C'' (black points) to a
    three-zone model (position in red, acceleration in orange).
    Descent starts with a free fall for a distance of
    $H_1\simeq28$~m and lasts $T_1=2.35$~s.
    In ensuing collision with the ground varying acceleration reveals
    that the top section consists of two zones, the primary
    below $H_2\simeq68$~m, for which $a\simeq6$~m/s$^2$, and
    the secondary above, for which $a\simeq-1$~m/s$^2$.
}
\end{figure}

Data set ``C,'' as presented to us, consists of two descent curves, one
at 10 samples per second, and the other at 5 samples per second.
We leave presentation and discussion of the descent curve to future publication
by their author.~\cite{wtc7:c}
In what follows we use low resolution data set only, which consists
of 25 data points, with an estimated error in distance of $\pm0.2$~m.

The importance of the physical model \noeqn{eq:eom:fema} is that it allows us
to identify the phases of collapse
(stages of descent between which the acceleration changes discontinuously)
to the zones of the top section being destroyed in the collision with the ground.

As a measure of how close is a solution of \eqn{eq:eom:fema},
call it $Z=Z(t)$, to the data set ``C'' we use a
sum-of-absolute-errors (SAE),
\begin{equation}
  \label{eq:sae}
  \mbox{SAE} =
  \sum_{i=1}^N \, \left|
      Z(t_i;\{r_j,s_j\}_j) - z_i
    \right| .
\end{equation}
Inspection of data suggests that there are three zones
in the building, each with its own $r$ and $s$.

We estimate $\{(r_i,s_i)\}_{i=1,3}$ as follows.
We perform a sequential fitting where in the first step we fix
heights $H_1$ and $H_2$ while varying simultaneously the
parameters of the resistive force in all three zones.
In the second step we vary the positions of one of the bounds, $H_1$ or
$H_2$, in 0.5 and 1~m increments, respectively, where we use the
results obtained in the previous step as the initial conditions.
For minimization of SAE we use a simplex method of Nelder and Mead,
as it does not require computation of derivatives.
For solving the ODE we use a Runge-Kutta Prince-Dormand (8,9) method.
Both methods are implemented in the GSL.\footnote{%
M. Galassi and J. Davies and J. Theiler and B. Gough and
G. Jungman and M. Booth and F. Rossi,
GNU Scientific Library, Reference Manual, Version 1.3, 2002,
{\tt http://www.gnu.org/software/gsl}.}

The results of the minimization are shown in \fig{fig:3}.
In the three-zone model the best fit achieves $\mbox{SAE}\simeq3.824$~m.
The boundaries between the zones are at $H_1=28$~m and $H_2=68$~m,
where $(r_1,s_1)=(0,0)$, $(r_2,s_2)=(0,0.41)$ and $(r_3,s_3)=(0,1.09)$,
for Phases I, II, and III of descent, respectively.
The fit of the model to the data is excellent with an average error
per point being $\sim0.15$~m, thus smaller then the
$\sim0.2$~m margin given by the set ``C'' author.
We estimate the width of the boundaries by introducing a number of
``micro'' zones at $H_1$ and $H_2$.
This procedure yields for the width of transition at $H_1$ of 1~m  and
2~m at $H_2$, which we write as $H_1 = 28\pm0.5$~m and $H_2=68\pm1$~m.
The accelerations during the descent are $a\simeq g$ for the free
fall phase (Phase I), $a\simeq6$~m/s$^2$ for the ``crush-up'' of
the primary zone (Phase II) and $a\simeq-1$~m/s$^2$ for the ``crush-up''
of the secondary zone (Phase III).

\begin{figure}[tph]
  \singlespacing
  \centering
  \includegraphics[scale=0.8,clip]{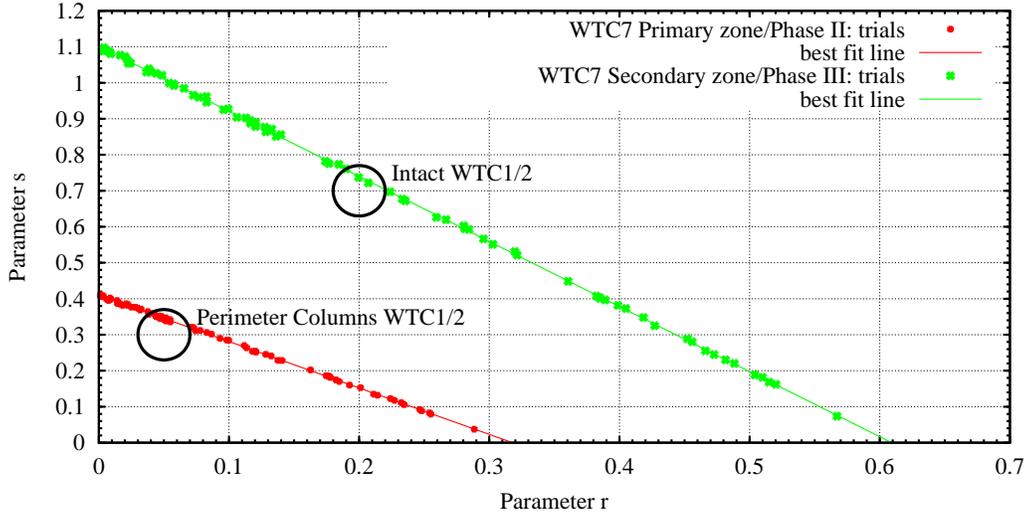}
  \caption{%
    \label{fig:4}
    Uncertainties in $(r,s)$ in the three-zone model obtained by
    allowing SAE, \eqn{eq:sae}, to vary up to 1\% from the minimum.
    As a result $(r_i,s_i)$, $i=2,3$, disperse along two straight lines,
    \eqn{eq:rs2:line} for Phase II, and \eqn{eq:rs3:line}
    for Phase III.
    At the same time, the duration of collapse spreads uniformly
    between 7.5 and 8.4~seconds.
    For comparison we also show our estimates for $r$ and $s$ in WTC 1 and
    2 (black circles) from~\cite{beck2006}:
    intact building with $(r,s)=(0.2,0.7)$, and the contribution from
    its perimeter columns which $(r,s)=(0.05,0.3)$.
    The radii represent their uncertainties: $\sim20\%$ for $r$, and
    half that for $s$.
    We see that the secondary zone of WTC 7 appears to be intact.
}
\end{figure}

We estimate uncertainties in $\{(r_i,s_i)\}_{i=1,3}$ as follows.
We perform a large number of optimizations $(N=100)$, where we
randomly choose initial values for $(r_j,s_j)$, while keeping
$H_1$ and $H_2$ fixed.
Further, of all so obtained $\{(r_i,s_i)\}_{i=1,3}$ we keep only
those which SAE is within 1\% of the best (smallest) value.
In~\fig{fig:4} we show the results of minimization.
While $(r_1,s_1)$ remain unremarkably close to 0,
in Phase II we find that $(r_2,s_2)$'s spread
along the line,
\begin{equation}
  \label{eq:rs2:line}
  \frac{r_2}{0.31} + \frac{s_2}{0.41}\simeq1.
\end{equation}
For Phase III we find that $(r_3,s_3)$'s spread along the line,
\begin{equation}
  \label{eq:rs3:line}
  \frac{r_3}{0.61} + \frac{s_3}{1.10}\simeq1.
\end{equation}
This indicates that SAE posses shallow minima along the lines
\eqn{eq:rs2:line} and \eqn{eq:rs3:line}.

We further reduce a number of zone parameters in the three-zone model by
considering that the top section throughout both zones is comprised of load bearing
elements of identical properties.
Then, the anticipated difference in strength between the zones comes
from varying their number of elements.
We introduce a constant $k$, which couples the resistive force in the primary
and the secondary zone, or in Phase II and III, respectively,
as follows
\begin{equation}
  \label{eq:phases23:con}
  (r_2,s_2) = k \cdot (r_3,s_3).
\end{equation}
While the results of optimization remain unremarkably close to the
one listed previously, this procedure yields a very narrow
estimate $k=0.42\pm0.04$.
In other words, the primary zone appears to be 60\%, or so, compromised
compared to the secondary zone.

\subsection{The Secondary Zone}

To our knowledge, there were no attempts to estimate the magnitude
of resistive force in WTC 7.
However, such estimates were provided for
WTC~2 by Ba\v{z}ant and Verdure~\cite{bazant2006}, and for WTC~1 and 2 by us~\cite{beck2006}.
The following discussion is based on an assumption that those estimates
represent reasonable values for WTC 7, as well:
\begin{itemize}
\item
  {\bf Ba\v{z}ant and Verdure's estimate:}
  The authors make an educated guess that the crushing energy $\Delta L$
  per floor in an intact building is $\Delta L = 0.6 - 2.4$ GNm.~\footnote{%
  In their numerical modeling they also allow arbitrary halving of their
  estimate of the resistive force.
  They argue that the reduction is due to increased temperature of the
  environment.}
  The resistive force is then given by $R = \Delta L / \Delta H$,
  where $\Delta H = 3.7$~m is the floor height.
  They take mass of WTC 1 and 2 to be $M=3.2\cdot10^{8}$~kg
  which yields $r = R/(M\,g) = 0.05-0.2$.
  The authors defend their estimate being so small by introducing a scenario of
  collapse which is currently being disputed.\footnote{%
  Most recent objections being raised by Szuladzinski, whose simulations of
  compression failure of individual columns suggest that their resistive
  force is much greater than Ba\v{z}ant and Verdure allow in their report.}
  Seffen~\cite{seffen2008} proposes an alternative, more obscure,
  catastrophic mechanism, which net effect is the same: a near-zero
  resistive force in a collapse of otherwise an intact building.

  Given the magnitude of the resistive force in three phases of collapse,
  the two catastrophic mechanisms can, at best, be applied to the rapid reduction
  of the base's strength to zero in Phase I.
  We note that the physical processes behind both mechanisms
  require the top section to produce ``destruction waves'' at the collision
  plane with the base.
  The ``waves'' propagate through the structure in front of the top section
  and reduce its resistive force to near-zero,
  after which the top section collects the pieces in its path.
  However, this cannot explain a free fall seen in WTC 7: the collected mass
  acts as an inertial brake which prevents the top section from
  ever reaching the free fall acceleration.
  We quantify this analysis in the next section.

  We conclude that their physical mechanisms, and consequently their estimates
  of resistive force, do not apply to WTC 7.
  Whether they apply to WTC 1 and 2 is yet to be seen.
\item
  {\bf Beck's estimate:}
  we argued that it suffices to consider
  a textbook model of resistive force,\footnote{%
  Please note, Ba\v{z}ant and Verdure ({\em ibid.}) and Seffen({\em ibid.})
  recognize the textbook model as an upper limit for the resistive
  force.
  Considering that {\em the experiment is the mother of all theories}
  it would be beneficial if members of scientific community would take
  upon themselves to examine the video footage of the collapse and
  measure or verify the descent curves for all three buildings.
  To this date the descent curves for WTC 1 and 2 are unavailable,
  and the NIST investigators ignore them in their investigation.
  } $R = \epsilon \cdot Y$, where $Y = Y(Z)$ is
  the ultimate yield strength of the vertical columns as a function of
  drop from the building's top,
  while $\epsilon\simeq 0.25$ is the ultimate yield strain of the
  structural steel used in the building.
  Using the estimates for properties of the vertical columns in
  WTC 1 and 2 (their cross section, strength and the weight they carry)
  one arrives to $Y/(M\cdot g) \simeq 0.8 + 2.7 \cdot Z/H$,
  that is, $(r,s) \simeq (0.2,0.7)$ in an intact building.
  On the other hand, the parameters of the secondary zone of WTC 7 are
  on the line $r_3/0.6+s_3/1.1 \simeq 1$.
  We see that, within a margin of error of under 5\%, the two overlap.
  We base our error estimate on the ratio of a distance between the
  two ($\sim0.04$) to $r+s=0.9$, and which is under 0.05.
  This is illustrated in~\fig{fig:4} where we plot $r$ and $s$ in
  the primary and the secondary zone in WTC 7 and compare it to their
  values in intact WTC 1 and 2, and to the contribution from perimeter
  columns only.
\end{itemize}

We conclude that the secondary zone of WTC 7 appears to be intact.

\subsection{The Primary Zone and the NIST Hypothesis Regarding the Collapse
Initiation}

\begin{figure}[tph]
  \singlespacing
  \centering
  \includegraphics[scale=0.8,clip]{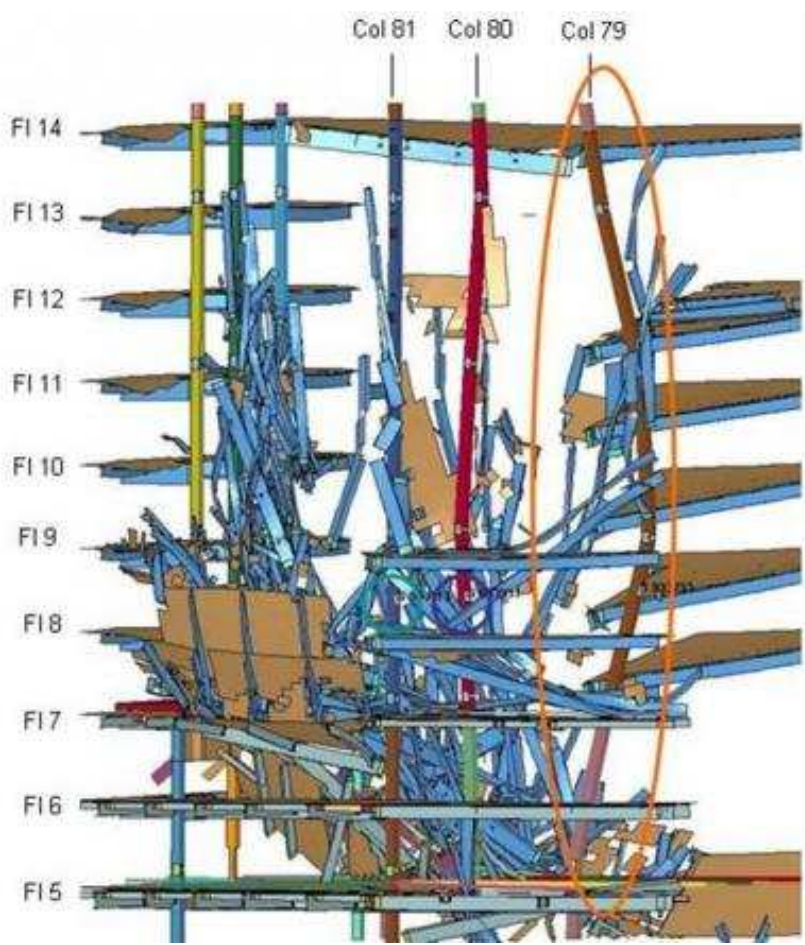}
  \caption{%
    \label{fig:nist}
    Status of the WTC 7 primary zone at the onset of collapse as proposed
    by the NIST investigators.
    The failure of column No. 79 near the 13th floor is presumably responsible
    for the initiation of the collapse.~\cite{nist:2008:wtc7}
}
\end{figure}

The descent curve provides us with a precise estimate of the primary
zone being 60\%, or so, damaged compared to the intact secondary zone.
Here, the primary zone stretches from $H_1=28$~m (4th floor)
to $H=68$~m (15th floor).
We use the findings by the NIST investigators to posit that the core columns
are absent in the primary zone.

We start by stating the hypothesis of the NIST investigators,~\cite{nist:2008:wtc7}
regarding the distribution of damage prior to the collapse
and its initiation, both illustrated in~\fig{fig:nist}.
The report blames a failure of the column No.~79 near the 13th floor
for the collapse initiation.
Its failure presumably induces a cascade of floor failures,
which cause the buckling of ``additional columns'' and ``within seconds,
the entire building core is failing.''

We immediately note that the last statement in their hypothetical failure
scenario is not correct:
according to the descent curve it is not the entire building core that
is failing but only the core below $H_2$.
We recall that the descent curve indicates that the secondary zone (the building
above $H_2$) managed to stay intact not just at the initiation of collapse
but also until the collision plane of destruction reached it some 4 seconds
into the collapse.
Now, if it is only the core below $H_2$ that is failing, than
this failure for sure includes the core columns of the primary zone,
and possibly the core columns of the base.
This is only the first of disagreements between the NIST hypothesis and
the descent curve.

Second, it is not clear what drives the failure of the core below $H_2$.
It is not the weight of the building above: the core columns are severed at
$H_2$ so below $H_2$ they carry little more then their own weight.
Thus, the mass participating in the ``cascade of floor failures'' at
best corresponds to the mass of the core below $H_2$.

Third, a connection between the cascade of floor failures and the
release of the top section at $H_1\simeq28$~m is not clear in the findings
by the NIST investigators.
The hypothetical failure that starts near $H_2$ propagates downwards through
the center, possibly to the  ground level, on one hand side.
On the other, at $H_1$ this ``cascade'' surfaces at the
sides of the building and severs the perimeter columns so that the
free fall may commence.
If the ``cascade'' indeed surfaces near $H_1$, and is ``spontaneous,''
then $H_1$ should vary considerably at the exit points
along the perimeter of the building.
In terms of the descent curve this would manifest itself as $H_1$
being a range over which the resistive force changes from zero to
some other value.
Contrary to being ``spontaneous,'' the descent curve indicates a sudden
transition from Phase I to Phase II, where its width is
less than 1~m.

We conclude that the NIST report, as is, agrees with the descent curve only
in regard to the distribution of damage in the primary zone: in a cascading
floor failure that started at the top of the primary zone the building,
most likely, lost all of its core columns in the primary zone and in
the base.
We emphasize this point in~\fig{fig:4} which, among others,
shows that $r$ and $s$ in the primary zone of WTC 7 are rather close
to the estimated contribution from the perimeter columns in WTC 1 and 2,
we presented in~\cite{beck2006}.

The NIST investigators, on the other hand, ignore the demise of the base
and a role it played in initiation of the collapse.

\subsection{Free Fall and the Demise of the Base}

\begin{figure}[tph]
  \singlespacing
  \centering
  \includegraphics[scale=0.8,clip]{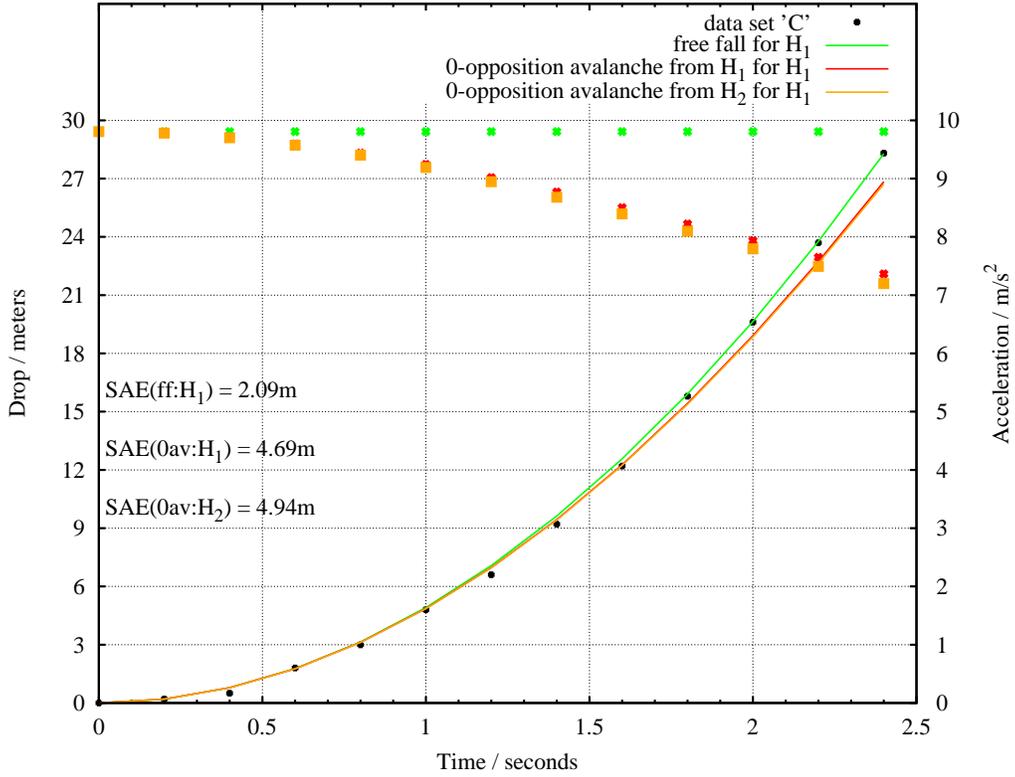}
  \caption{%
    \label{fig:base:freefall}
    (Phase I) Trajectories for the theoretical models, free
    fall (green, solid line for position, dots for acceleration),
    an 0-opposition avalanche from $H_1=28$~m to the ground level (red),
    an 0-opposition avalanche from $H_2=68$~m (orange)
    and the actual descent curve from the data set ``C'' (black dots).
    Also given is the SAE between the descent curve and a model.
    In terms of SAE, the free fall is the best fit to the descent curve.
    SAE over Phase I for either avalanche is greater than that for the
    three-zone model for an entire descent ($\simeq3.84$~m).
    Furthermore, discrepancies between the free fall and the descent curve are the greatest at
    the first 1.5~s, during which the descent data is known to be imprecise
    due to a low resolution of the recorded video.
    Conversely, the avalanches show systematic departure from the descent curve
    in the second half of Phase I, where the descent curve is (relatively) more
    precise.
}
\end{figure}

From the beginning it transpired that the top section flew
through the base in a fashion awfully close to a free fall.
Later when discussing the secondary zone we mentioned an alternative
physical model of descent: an actual avalanche.
We now examine more closely Phase I of the descent for a distance
$H_1\simeq28$~m as described by the data set ``C'' and compare
it to the free fall motion, and to a motion of a fictional avalanche
that started at some height in the building, say, $H^*$ and continued for $H_1$.

Here we recall that a difference between the avalanche and the free fall
is that in the avalanche it is the top section that destroys a part of the
base in its path and then adsorbs it,
while in the free fall the base is quickly converted into the free floating chunks so
the entire building simultaneously free falls.
Also, as discussed earlier,
Ba\v{z}ant and Verdure~\cite{bazant2006}, and Seffen~\cite{seffen2008}
proposed a highly speculative mechanisms which net-result is that
the avalanche by the top section feels almost-zero resistance when crushing,
presumably, intact structure of the base in its path.
NIST investigators imply the same in the quotation we stated earlier.

We start by stating the equation of motion of a 0-opposition
avalanche that at time $t$ has dropped to position $Z$,
\begin{equation}
  \label{eq:avalanche:eom}
  \frac{d}{dt} \left( Z\dot Z \right) = g \, Z,
\end{equation}
where for simplicity we neglect the effects of compaction $\kappa$.
Assuming that the motion at $t=0$ starts from rest, $\dot Z(0) = 0$,
at drop $Z(0)=Z_0$,
\eqn{eq:avalanche:eom} can be integrated once, yielding a relationship
between the current velocity and the position.
The time $t$ it takes the avalanche to propagate from $Z_0$ to $Z_1$
is given by
\begin{equation}
  \label{eq:avalanche:time}
  t(Z_0,Z_1) =
  \int_{Z_0}^{Z_1} \frac{dZ}{\dot Z}
  = - 1.82 \, \sqrt{\frac{Z_0}{g}}
  + 2.45 \, \sqrt{\frac{Z_1}{g}} \cdot
  {_2 F _1} \left( -\frac 1 6, \frac 1 2; \frac 5 6;
    \frac{Z_0^3}{Z_1^3} \right),
\end{equation}
where $_2 F _1 = {_2 F _1}(a,b;c;z)$ is the hyper-geometric
function~\cite{GradshteynRyzhik}.
Similarly, the acceleration as a function of position for $Z\in[Z_0,Z_1]$
is given by
\begin{equation}
  \label{eq:avalanche:acc}
  \ddot Z = \frac{g}{3}\,\left(
    1 + 2\, \frac{Z_0^3}{Z^3} \right).
\end{equation}
Finally, we use \eqn{eq:avalanche:time} and \eqn{eq:avalanche:acc}
to find the drop of the avalanche and its acceleration at the
times at which data set ``C'' was recorded.
For completeness we consider two avalanches, the first, that starts
at $Z_1 = H - H_1$ and propagates to the ground level, $Z_2=H$,
and the second, that starts at $Z_1 = H-H_2$ and propagates for $H_1$
to $Z_2=H-H_2+H_1$.

In \fig{fig:base:freefall} we show the data set ``C,'' the
respective theoretical trajectories, and their accelerations and SAE.
Here, SAE is calculated over the first 13 points of the data set ``C'' and
the positions predicted by each theoretical model.
We note that the free fall has SAE$\simeq2.09$~m, on one hand side,
and on the other, that SAE for the entire three-zone crush-up model
over 25 points ($\simeq3.82$~m) is less than
SAE for either of the 0-opposition avalanche models ($\simeq4.69, 4.94$~m).
This finding reaffirms our previous conclusion that Phase I is a free fall
for $H_1\sim28$~m and not an avalanche that started somewhere in
the building and propagated for the same distance.

In light of our discussion of the secondary and the primary zone we speculate
what must happen to the base for a free fall of the top
section to be possible:
\begin{itemize}
\item[1.]
  The core of the base is destroyed in the same sequence
  as the core of the primary zone.
  The NIST investigators appear to be hypothesizing this to be the case:
  what they believe is a cascade of floor failures  may in fact be
  a staged destruction (severing) of the core columns in the
  primary zone and then continues throughout the base.
  However, the damage to the base is more extensive than the damage to the
  primary zone in that in the base
  the floors and their web of trusses are destroyed as well.
  The top section is later, at $t=0$, released by severing
  the perimeter columns at $H_1$.
  These columns offer little to no resistance to the falling top section
  due to their marginal position and small cross section.
\item[2.]
  The destruction of the entire base at $t=0$ is inconsequential to
  an earlier destruction of the core columns in the primary zone.
  The strength, of otherwise intact, base is sufficient to arrest
  the fallout of staged destruction of the core columns in the primary
  zone.
  Here, the base being annihilated is what releases the top section.
\end{itemize}
We note that in terms of the apparent weight $W$ the building exerts on the
Earth's crust during its collapse, \eqn{eq:weight},
the two cases differ.
There are two type of terms contributing to the apparent weight:
``arrest,'' created by a large chunk of the building coming to a stop
after hitting the ground or a part of the building in its path,
and ``release,'' created by a large chunk breaking off the building
and starting a free fall.
It is a reasonable assumption that the seismic signal is
excited by changes in the apparent weight of the building, $\delta w$,
given by $\delta w = f \, \hat\Delta W / (M\,g)$, where
$f$ is the sampling rate, while $\hat\Delta$ is a difference operator
acting on a time series of $W$ collected at the sampling rate.

It can be shown that if the interior, core and floors, of the base
and the core of the primary zone
are destroyed prior to the release of the top section (Case 1) than
the peak in $\delta w$ from the first release (the top section
being allowed to free fall) is comparable to,
possibly weaker than, the peak of the
first arrest (the top section reaching the ground).
On the other hand, if the destruction of the entire base marks the release
of the top section (Case 2) then the peak of the first release
is much stronger than the peak of the first arrest.

We believe it is the seismic signal of the collapse that can be used to deduce
which of the two cases is more likely to have had occurred.
We leave this analysis to future publications with our collaborators.
Given our current knowledge~\cite{seismo}, we favor Case 1.

\newpage
\section{Conclusion}

\begin{figure}[tph]
  \singlespacing
  \centering
  \includegraphics[scale=0.6,clip]{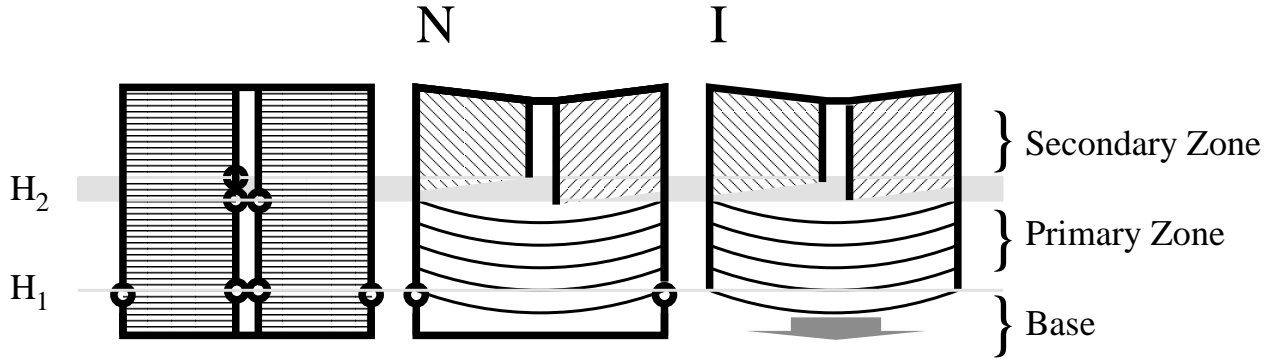}
  \caption{%
    \label{fig:collapse}
    Not-to-scale illustration of the first few phases of collapse of WTC 7.
    In the initial building, left panel, during Preparatory
    or Null Phase the core columns are destroyed between the heights
    $H_2=68$~m and $H\simeq28$~m, middle panel,
    which splits the top section in two: the 60\% damaged primary, and
    the intact secondary zone.
    The interior of the base is destroyed, as well.
    Phase I, right panel, starts when the perimeter columns
    in the base are severed at $H_1$, allowing the top section
    to free fall. Phase II, the ``crush-up,''
    begins when the top section reaches the ground (not shown).
}
\end{figure}

Given the descent curves, the results of physical modeling,
some video evidence and the damage review by the NIST investigators
we conclude that the collapse of WTC 7 is comprised of four phases:
\begin{itemize}
\item
  {\bf Phase N}: Null or Preparatory phase starts 8, or so, seconds
  before the collapse.
  During that phase, we argue, the core between $H_2\simeq68$~m (15th floor)
  and $H_1\simeq28$~m is destroyed together with the base interior.

  The appearance of the building during that period, which features,
  among others, sinking of the penthouses on the top into the building,
  is consistent with severing of the core columns below $H_2$.
  The sinking results from the sections of the core columns above $H_2$
  being left suspended from the hat truss and the perimeter
  columns.
  That these hanging sections of core columns in the secondary zone
  are not destroyed becomes apparent during Phase III when the top
  section in its last moments regains its full (local) strength.
\item
  {\bf Phase I}: Free Fall phase begins at $t=0$ with a sudden and total
  annihilation of the base (part of the building between the ground level
  and $H_1$).
  This allows the top section (part of the building above $H_1$)
  to free fall to the ground.
\item
  {\bf Phase II}: ``Crush-up'' begins $t \simeq 2.3$~s into the collapse
  when the top section reaches the ground.
  For the next $\sim$42~m the primary zone of the top section, which
  was compromised during Phase N, is destroyed in collision with the ground.
\item
  {\bf Phase III}: ``Crush-up'' of the top section continues for the next
  $t \simeq 3.8$~s as the secondary zone is being destroyed.
  While the top section now begins to decelerate, this, in itself,
  is not sufficient to arrest the collapse.
  The phase continues some 7.8-8.8~s into the collapse when the last remains
  of the building fall on the ground.
\end{itemize}

We conclude that the building was destroyed in a highly controlled
fashion and, contrary to the common sentiment,
did not spontaneously collapse.

\begin{acknowledgements}
  This report was first published on-line on {\tt www.arxiv.org} and
  was based on, to that date available, descent curve ``N'' published
  by an organization {\em 9-11 Research}.
  It motivated Mr. JPA to contact us in November of 2008, and through
  his contacts provide us with to date most extensive descent curve ``C.''
  Without curve ``C'' by Mr. David Chandler, the main findings of this
  report - free fall and a presence of two zones in the top section, one
  of which is intact -
  would be highly speculative, as readers can verify from the earlier versions of
  this report.

  We acknowledge stimulating discussions with Mr. JAP, Mr. Chandler, and Mr. E,
  and their critical remarks.
  We acknowledge the feedback provided by the members of different public
  Internet forums and their critical remarks.
\end{acknowledgements}

\newpage
\bibliography{%
  bibliography/applied%
}

\appendix
\newpage
\section{Data Set ``C''}
\begin{table}[h]
  \caption{
    \label{tab:c}
    Raw data comprising the descent curve ``C.''\cite{wtc7:c}
    It measures a height of a point on the building, which exact
    position is identified from the NIST report.
}
  \begin{tabular}{|c|c|}
    \hline\hline
    Time  & Drop\\
    (sec) & (m)\\
    \hline\hline
    2.0&  155.80\\
    2.2&  155.60\footnote{%
      We use value 155.80~m instead, and posit that
      the uncertainty in the drop is $\pm0.2$~m.
      Only effect of this action is that SAE is reduced
      by 0.2~m by hand.
}\\
    2.4&  155.60\\
    2.6&  155.30\\
    2.8&  154.00\\
    3.0&  152.80\\
    \hline
    3.2&  151.00\\
    3.4&  149.20\\
    3.6&  146.60\\
    3.8&  143.60\\
    4.0&  140.00\\
    \hline
    4.2&  136.20\\
    4.4&  132.10\\
    4.6&  127.50\\
    4.8&  122.90\\
    5.0&  117.80\\
    \hline
    5.2&  112.70\\
    5.4&  107.10\\
    5.6&  101.80\\
    5.8&   95.38\\
    6.0&   89.52\\
    \hline
    6.2&   82.63\\
    6.4&   76.77\\
    6.6&   70.14\\
    6.8&   64.52\\
    7.0&   58.15\\
    \hline\hline
  \end{tabular}
\end{table}

\end{document}